\documentclass[12pt,a4]{article}
\usepackage{graphicx}
\usepackage{longtable}
\usepackage{amsmath}
\usepackage[all]{xy}

\newcommand{\be}{\begin{equation}}
\newcommand{\ee}{\end{equation}}
\newcommand{\ba}{\begin{eqnarray}}
\newcommand{\ea}{\end{eqnarray}}
\newcommand{\baa}{\begin{eqnarray*}}
\newcommand{\eaa}{\end{eqnarray*}}

\renewcommand{\k}{{\vec k}}

\begin{document}

\title{Dirac point metamorphosis from third-neighbor couplings in graphene}
\author{Cristina Bena$^{1,2}$, Laurent Simon$^3$\\
{\small \it $^1$Laboratoire de Physique des Solides, Universit\'e
Paris-Sud},
\vspace{-.1in}\\{\small \it  B\^at.~510, 91405 Orsay, France}\\
{\small \it $^2$Institut de Physique Th\'eorique, CEA/Saclay,
CNRS, URA 2306},
\vspace{-.1in}\\{\small \it  Orme des Merisiers, F-91191 Gif-sur-Yvette, France}\\
{\small \it $^3$ Institut de Sciences des Mat\'{e}riaux de
Mulhouse IS2M-LRC 7228-CNRS-UHA}, \vspace{-.1in}\\{\small \it 4,
rue des fr\`eres Lumi\`ere 68093 Mulhouse-France}} \maketitle
\begin{abstract}
We study the band structure and the density of states of graphene in the presence of a next-to-nearest-neighbor coupling (N2) and a third-nearest-neighbor coupling (N3). We show that for values of N3 larger or equal to $1/3$ of the value
of the nearest-neighbor hopping (NN), extra Dirac points appear in the spectrum. If N3 is exactly equal to $1/3$ NN, the new Dirac points are localized at the M points of the Brillouin zone and are hybrid: the electrons have a linear dispersion along the $\Gamma M$ direction and a quadratic dispersion along the perpendicular direction $MK$. For larger values of N3 the new points have a linear dispersion, and are situated along the $MK$ line. For a value of N3 equal to $1/2$ NN, these points merge with the Dirac cones at the K points, yielding a gapless quadratic dispersion around K, while for larger values each quadratic point at K splits again into four Dirac points. The effects of changing the N2 coupling are not so dramatic. We calculate the density of states and we show that increasing the N3 coupling lowers the energy of the Van Hove singularities, and when N3 is larger than $1/3$ NN the Van Hove singularities split in two, giving rise to extra singularities at low energies.
\end{abstract}

\section{Introduction}
Graphene and graphene-like materials have received an increasing
amount of interest over the past years, in particular because of
the exotic physics arising from its Dirac-like, linear
excitations. Some important questions in the study of graphene are what is the mechanism giving rise to Dirac points in the spectrum, in what circumstances their number
and their position can be modified, as well as under what
conditions the linear Dirac-like quasiparticles acquire a
quadratic dispersion and possibly become gapped. Recent works
\cite{jap,gilles1,gilles2,castroneto} have proposed that by
modifying one of the three couplings between a carbon atom and its neighbors, one can modify the position of the Dirac points in the spectrum. Moreover, for a critical
value of the hopping asymmetry, each pair of Dirac points will
merge forming a hybrid point in whose vicinity the quasiparticles
have a Dirac-like dispersion along one direction, and a quadratic
dispersion along the other. The existence of this peculiar type of nodal points gives rise to various unconventional effects such as an atypical dependence of the Landau level energies on magnetic field
\cite{gilles1}. Furthermore, such an anisotropy in the NN hopping energy has been found to introduce strong modifications of the Klein tunneling \cite{TreidelPRL2010}, and also to give rise to a phase transition between gapped massive fermions and massless Dirac fermions \cite{ZhuPRL2007}.

Such modifications of the interatomic coupling
constants could be generated by applying uniaxial pressure, shear or strain forces on a graphene layer
\cite{castroneto,gapgraphene}. However, in order to have significant Dirac-point
displacements as well as merging and gapping of Dirac points,
the values of the required applied forces are quite high and at
the limit of the present experimental reach. Also the natural ripples of graphene sheets (which can also be associated to Peierls distortions) generate hopping energy fluctuations proportional to the bond length distortion \cite{Zhu2010arXiv}. However the out-of-plane deformation expected to have a significant effect on the hopping energy is expected to be high (up to 0.8 nm).

Here we propose a novel mechanism to change the number of Dirac points in graphene and engineer their positions and their properties. Our proposal is
based on the modification of the values of
the next-to-nearest neighbor coupling (N2), and in particular of the third-nearest-neighbor coupling (N3). It has been shown that the
N2 and N3, while generally ignored, are not
insignificant \cite{rotenberg} and may substantially affect the
band structure, in particular at high energies, close to the Van
Hove (VH) singularities. The cited values of N2 and N3 are of the order of $0.1 eV$ to $1 eV$ (N2) and $0.3 eV$ to $0.45 eV$  (N3), while the nearest-neighbor hopping (NN) is of order of $2.5$ to $ 3 eV$ \cite{rotenberg,tb1}. We will discuss possible experimental routes towards modifying the values of these couplings in section 6.

In particular, we show that for N3 equal to $1/3$ NN, new Dirac points arise at the M points. Exactly at $1/3$ the new Dirac points are hybrid,
such that the dispersion of the quasiparticles is linear along
the $\Gamma M$ direction, but quadratic in the $MK$ direction (see Fig.~1 for a definition of the crystallographic points). For larger N3, each new Dirac point splits into two linear Dirac points that move along the $MK$ direction, and
approach the K points with increasing the N3 coupling. For
N3 equal to $1/2$ NN, each Dirac point and
its three satellites merge to form gapless quadratically dispersing features around the K points, similar to the ones in bilayer graphene. For even larger values of N3, these features split giving rise each to four standard Dirac points.
These Dirac points are situated along $\Gamma K$ and approach the point $L$ (with $\Gamma L=1/2 \Gamma K$) for extremely large values of the N3 coupling. This is indeed expectable when N3 dominates as the periodicity of the lattice in this limit is effectively doubled.

The effect of the N2 coupling, on the other hand, is less
spectacular. Its main effects are an
asymmetry between the conduction and the valence band, a displacement
of the energy of the Dirac points (an effective doping), as well an inclination of the Dirac cones.

We also calculate the density of states of this system, and we find that increasing the N3 coupling shifts the energy of the Van Hove (VH) singularities towards lower energies. Moreover, for large enough values of N3, the VH singularities split, giving rise each to two singularities among which the one lowest in energy is in an experimentally accessible range.
As noted also in \cite{andrei}, being able to explore a system in the vicinity of a VH singularity may provide the tools to understand phenomena such as charge density waves, and high temperature superconductivity.

\section{Model}
The model we consider is a tight-binding model for graphene
\cite{tb1,tb2}: \be H=t \sum_{i,j=NN} c^{\dagger}_{i A} c_{j
B}+t'\sum_{i,j=N2}(c^{\dagger}_{i A} c_{j A}+c^{\dagger}_{i B}
c_{j B})+t''\sum_{i,j=N3}c^{\dagger}_{i A} c_{j B}+h.c. \ee In
momentum space this becomes \ba H=(\begin{array}{cc} c^\dagger_{\k
A} & c^{\dagger}_{\k B} \end{array} ) \left(\begin{array}{cc}
\epsilon_2(k_x,k_y)&\epsilon_1(k_x,k_y)+\epsilon_3(k_x,k_y)\\
\epsilon_1^*(k_x,k_y)+\epsilon_3^*(k_x,k_y)&\epsilon_2(k_x,k_y)\end{array}\right)
\left(\begin{array}{c}c_{\k A} \\c_{\k B}\end{array}\right)
\label{ht} \ea where \ba
\epsilon_1(k_x,k_y)&=&-t[e^{-i(k_x \sqrt{3}/2-k_y/2)}+e^{i(k_x \sqrt{3}/2+k_y/2)}+e^{-i k_y}]\nonumber \\
\epsilon_2(k_x,k_y)&=&-t'[e^{i(k_x \sqrt{3}/2 + 3
k_y/2)}+e^{i(-k_x \sqrt{3}/2 + 3 k_y/2)}+e^{i(k_x \sqrt{3}/2 -3
k_y/2)}+ \nonumber \\&& +e^{i(-k_x \sqrt{3}/2 -3 k_y/2)}+e^{i
\sqrt{3} k_x}+e^{-i \sqrt{3} k_x}] \nonumber \\
\epsilon_3(k_x,k_y)&=&-t''[e^{-i(k_x \sqrt{3}+k_y)}+e^{i(k_x \sqrt{3}-k_y)}+e^{2 i k_y}]
\ea Here $t$ is the NN
coupling, $t'$ the N2 coupling and $t''$ the N3 coupling (see
Fig.~1a)).

As described in many previous references \cite{review}, in the absence of N2 and
N3, the Hamiltonian can be linearized in the vicinity of the K
points of the Brillouin zone (BZ) which are usually denoted as
Dirac, or nodal points (see Fig.~1b). In Fig.~1b we plot the equal
energy contours in  the absence of N2 and N3 couplings, with the
usual linearly-dispersing quasiparticles being observed at the K
points.
\begin{figure}[htbp]
\begin{center}
\includegraphics[width=5in]{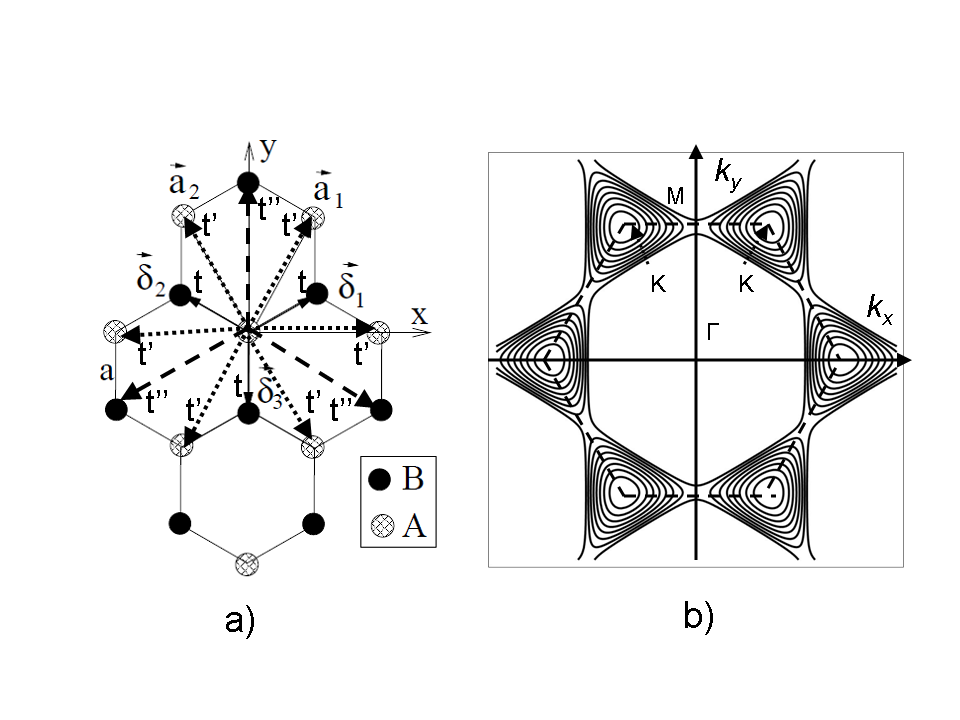}
\vspace{-0.3in} \caption{\small a) Graphene lattice, with NN, N2,
and N3 couplings indicated. b) Band structure of graphene in the
absence of N2 and N3 couplings. The Brillouin zone is marked by
dashed lines, and  the K, M and $\Gamma$ points are specified}
\label{fig1}
\end{center}
\end{figure}

The spectrum of the Hamiltonian in \ref{ht} is given by: \be
\epsilon(k_x,k_y)=\epsilon_2(k_x,k_y)\pm|\epsilon_1(k_x,k_y)+\epsilon_3(k_x,k_y)|
\ee
In what follows we will analyze how this spectrum evolves with increasing the N2 and N3 couplings.

\section{Increasing the N3 coupling}

We will first focus on how the energy spectrum evolves by the
introduction of an N3 coupling. We will take $t=3eV$ throughout our calculations.

\subsection{N3 of the same sign as NN}
For small values of $t''$ the modifications of the band structure are small (see Fig.~2), for example the equal-energy triangular contours in the vicinity of the Van Hove singularity become concave. This effect has already been observed at high energy in heavily doped graphene using ARPES \cite{rotenberg}. In Ref.\cite{rotenberg}, as soon
as the Van Hove singularity is approached with doping, the band becomes flat around the M points. Note also that the ARPES measurements display inhomogeneous intensity along the $KMK$ direction, whose origin is yet to be investigated.

Beside the deformation of the isoenergy contours, we can see that another interesting phenomenon arises: the slope of the bands in the vicinity of the K points is reduced such that the band energy at the M point is lower (see Fig.~2). Moreover, for values of $t''$ larger than $1/4t$, the band acquires an extra local minimum at the M point, and for a
critical value of $t''=1/3 t$, the energy of this point approaches zero, forming
a hybrid Dirac point, similar to the one described in
\cite{jap,gilles1,castroneto}.
\begin{figure}[htbp]
\begin{center}
\includegraphics[width=5in]{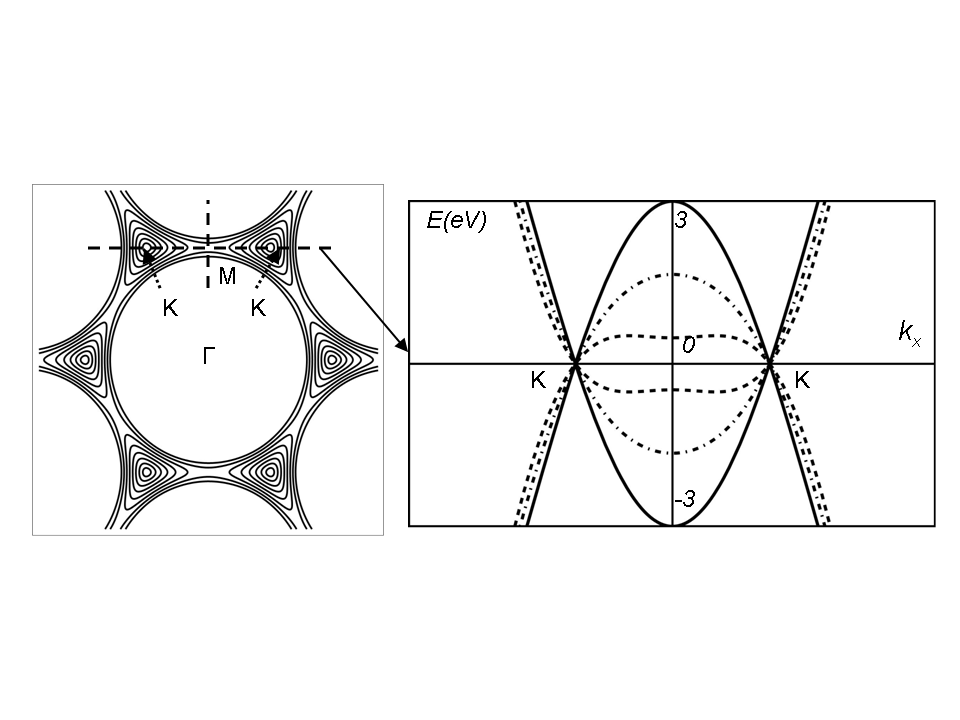}
\caption{\small a) Band structure of graphene for
$t''=0.15t=0.45eV$. Compared to $t''=0$, the triangles at the K points are rounded in, and the high
energy contours just above the VH transition are circular, instead
of hexagonal. b) Graphene energy dispersion for increasing values
of $t''$. The full line corresponds to $t''=0$, the dot-dashed
line to $t''=0.15t=0.45eV$ and the dashed line to
$t''=0.28t=0.84eV$. We can see that with increasing $t''$ the
bands curve more and more, such that the value of the band maximum
at M, and correspondingly, the position of the van Hove
singularity, are reduced. Moreover, for values of $t''$ larger
than $1/4t$, a local minimum appears in the spectrum} \label{fig2}
\end{center}
\end{figure}
The exact value of $t''$ for which a new Dirac point appears at M
can be obtained by performing an expansion of the band structure
close to one of the M points, for example $(0,2\pi/3)$:

\ba &&\epsilon(k_x,k_y)=\pm \bigg[(t-3t'')^2-\frac{3}{2}(t-4
t'')(t-3 t'')k_x^2+\frac{9}{16}(t-t'')^2 k_x^4+
\nonumber\\&&\frac{9}{2} (t^2+t t''-4 t''^2)
k_y^2-\frac{27}{16}(t^2+11 t t''-16t''^2)k_x^2
k_y^2\bigg]^{1/2}\nonumber \ea We note that this expansion
predicts that for a value of $t''$ bigger than $1/4$ the quadratic
curvature along the $x$ direction will change sign at M, giving
rise to a local minimum, as it can be seen in Fig.~2. Moreover, we
see that for $t''=1/3 t$, the energy goes to zero at the M point.
However, the new Dirac point arising for this particular value of
$t''$ is `hybrid' in that it exhibits a linear dispersion along
the $y$ axis, and a quadratic dispersion along the $x$ axis. This
is seen directly from the above analytical expansion for
$t''=1/3$:

\be \epsilon(k_x,k_y)=\frac{t}{4}\sqrt{64 k_y^2+k_x^4-78 k_x^2
k_y^2} \ee

A similar type of hybrid Dirac point has been predicted in
\cite{jap,gilles1,castroneto}, for graphene under pressure, for
which one of the three coupling between a carbon atom and its
three neighbors becomes twice as large as the other two.

\begin{figure}[htbp]
\begin{center}
\includegraphics[width=4in]{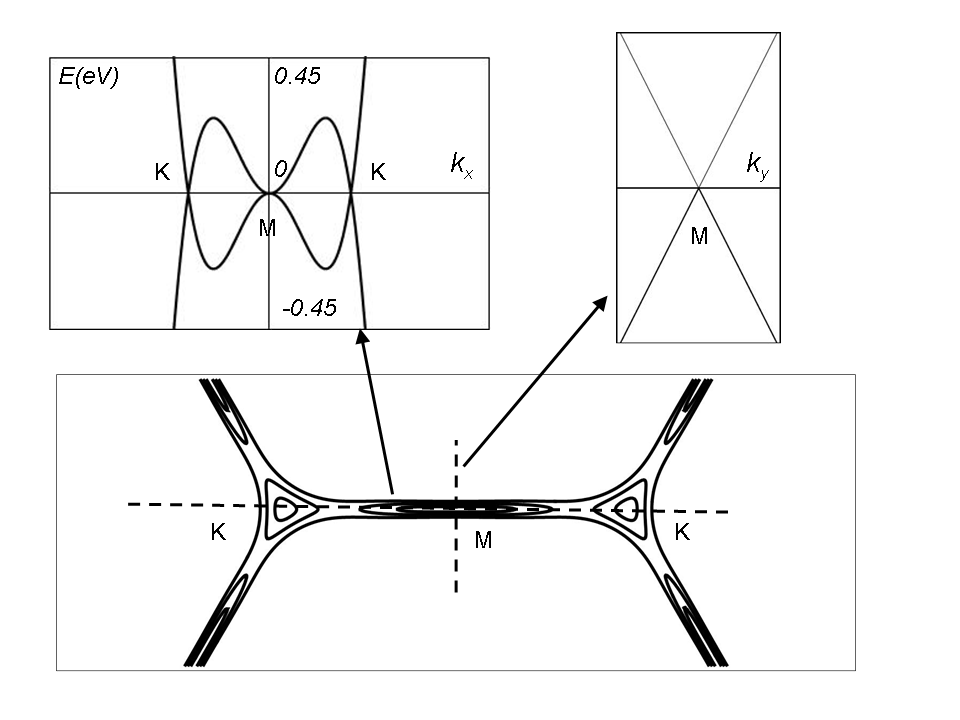}
\vspace{-0.1in} \caption{\small Graphene energy dispersion for $t''=1/3t$. We can see the appearance of a hybrid
Dirac point at the M point of the BZ} \label{fig1}
\end{center}
\end{figure}

If the N3 coupling is increased above the critical value of $1/3 t$, each
hybrid M point splits into two classical Dirac points that have
linear dispersion in all directions. These points
evolve towards the K points with increasing $t''$ (see Fig.~4). Due to the hexagonal symmetry of the system each K point will be surrounded by three satellite Dirac points.
\begin{figure}[htbp]
\begin{center}
\includegraphics[width=4in]{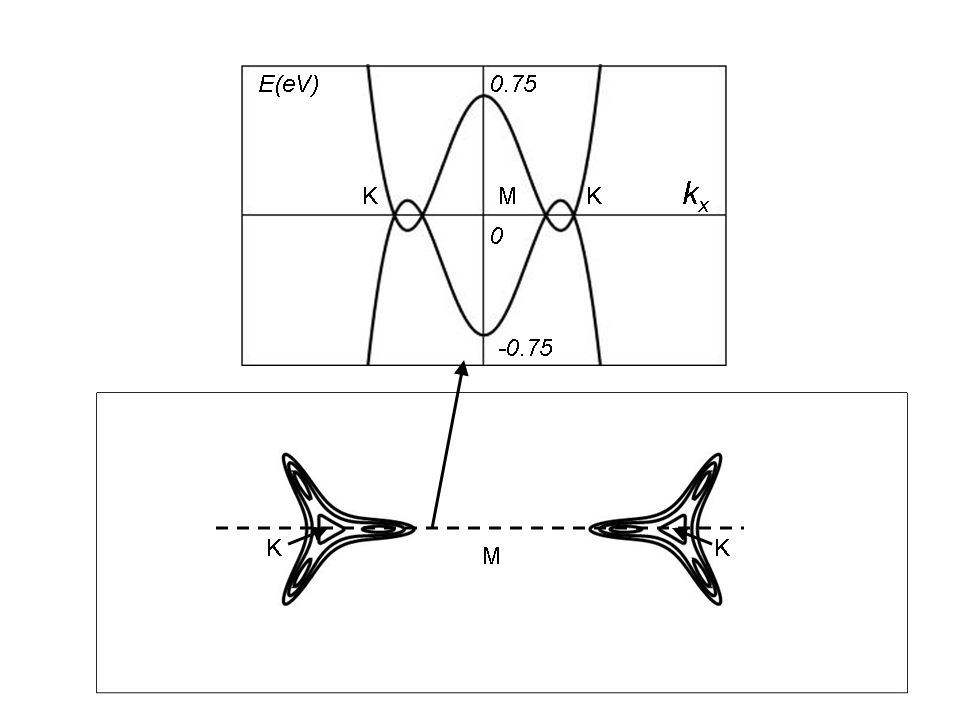}
\vspace{-0.2in} \caption{\small Graphene energy dispersion for
$t''=0.4 t$. We can see the appearance of two extra Dirac points
along the KMK axis that get closer and closer to K as $t''$
increases} \label{fig4}
\end{center}
\end{figure}
At a second critical value of $t''=1/2 t$, the satellite- and the K-Dirac points merge to form  gapless ``quadratic''
points (see Fig.~5), similar to the ones in bilayer
graphene. This can be seen from the expansion of the band
structure close to the K points \ba &&\epsilon(k_x,k_y)=\\&& =\pm
\bigg[\frac{9}{4} (t-2 t'')^2 k_y^2+\frac{9}{64} (t+4 t'')^2
k_y^4-\frac{27}{8} (t-2 t'')(t+4 t''^2) k_y^2 x \nonumber \\&&
+\frac{9}{4} (t-2 t'')^2+\frac{9}{32} (t+4 t'')^2 k_y^2
k_x^2+\frac{9}{8} (t-2t'')(t+4 t''^2) k_x^3+\frac{9}{64} (t+4
t'')^2 k_x^4\bigg]^{1/2} \nonumber \ea

We can see that away from
$t''=1/2 t$ the energy spectrum around the K point is given by:
\be \epsilon(k_x,k_y)=\pm \frac{3}{2} (t-2 t'') \sqrt{x^2+y^2} \ee
The non-zero value of $t''$ contributes to a reduction of the
Fermi velocity near the K point, which for $t''=0$ is equal to $3
t/2$. This is consistent with a reduction of the energy of the VH
singularity in the presence of N3 coupling, as described above,
as well in Ref.\cite{rotenberg}

Note that at the K point the spectrum is linear for all values of $t''$ except for $t''=1/2 t$, when the
dispersion in the vicinity of K becomes quadratic: \be
\epsilon(k_x,k_y)=\pm \frac{9}{8} t (x^2+y^2) \ee

\begin{figure}[htbp]
\begin{center}
\includegraphics[width=4in]{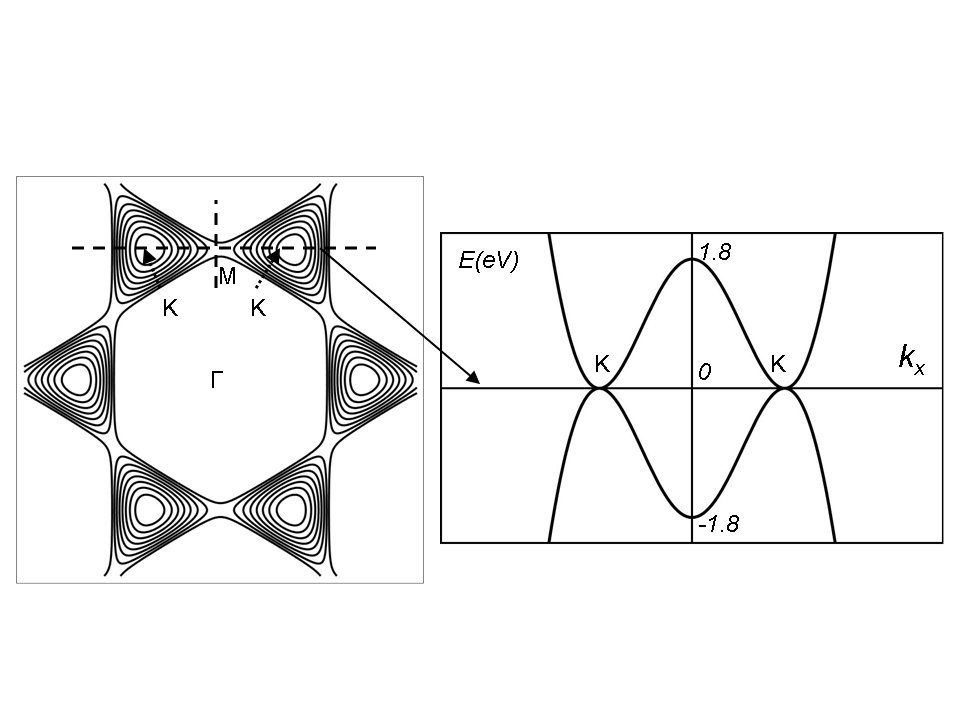}
\vspace{-0.3in} \caption{\small Graphene energy dispersion for the
particular case $t''=1/2 t$. We can see the appearance of
quadratic bands touching at the K points} \label{fig5}
\end{center}
\end{figure}

If the coupling $t''$ becomes larger than $t/2$, a new set of
three Dirac points is generated in the vicinity of each K point.
This time the Dirac points are situated along the $\Gamma K$ direction and along the two corresponding directions rotated by $\pm120^{\circ}$ with respect to $\Gamma K$ (see Fig.~6).

\begin{figure}[htbp]
\begin{center}
\includegraphics[width=4in]{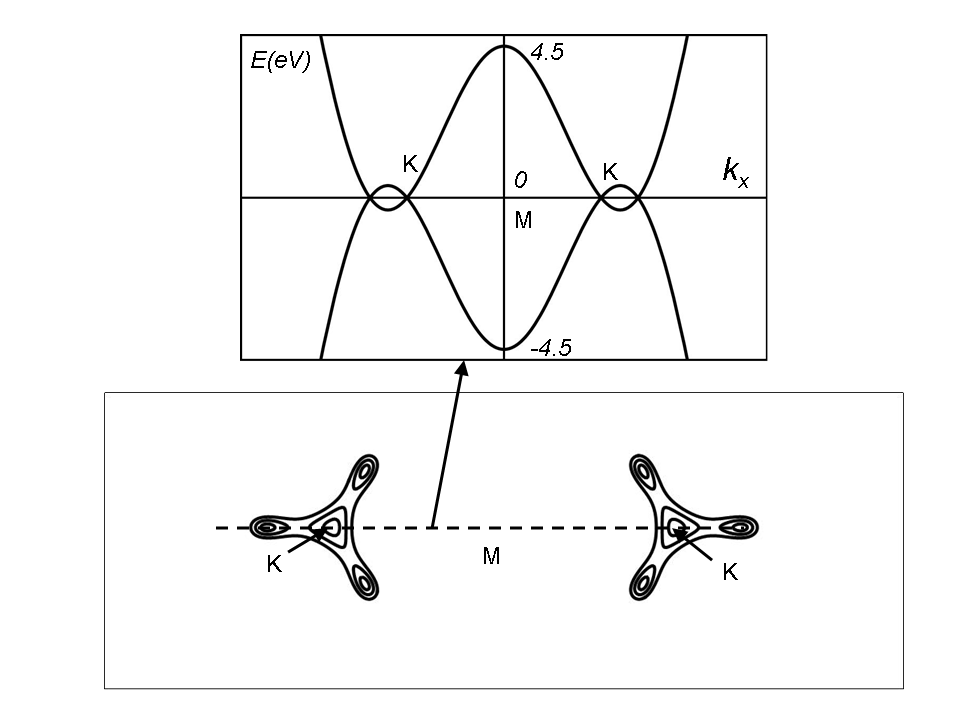}
\vspace{-0.2in} \caption{\small Graphene energy dispersion for
$t''=0.8 t$. New Dirac points develop along the direction $\Gamma
K$ and move towards the center of the BZ with increasing $t''$}
\label6
\end{center}
\end{figure}

When $t''$ dominates over $t$, the new Dirac points form
a k-space structure with a periodicity twice as large as the unperturbed one. This is expected, since a dominant N3 coupling is equivalent to having a lattice with a doubled interatomic distance (see Fig.~7).
\begin{figure}[htbp]
\begin{center}
\includegraphics[width=4in]{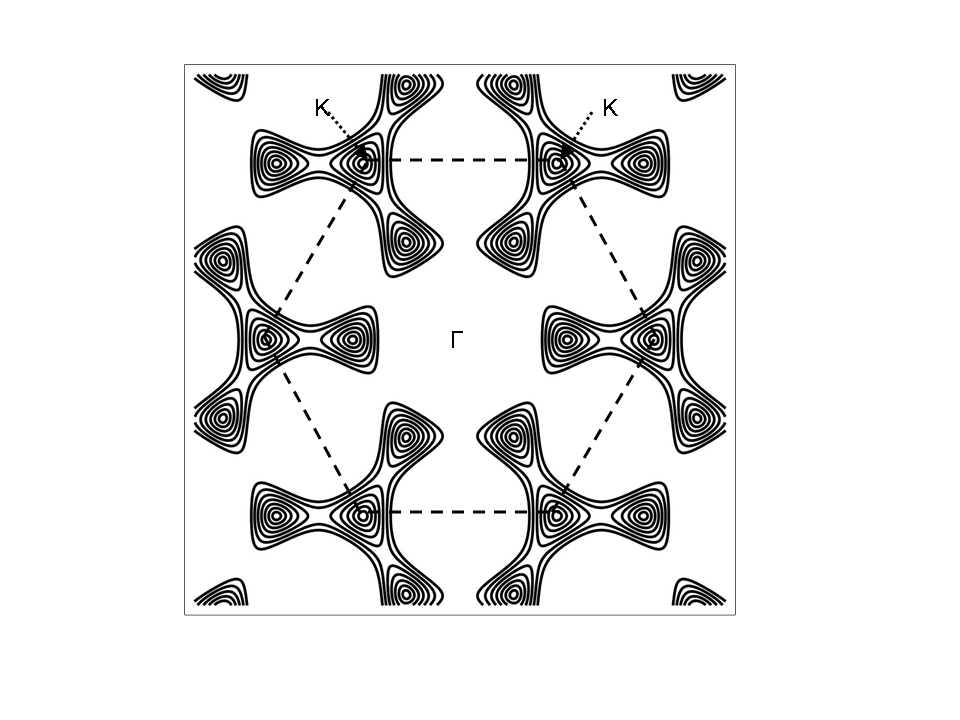}
\vspace{-0.3in} \caption{\small Graphene energy dispersion for
$t''=5 t$.} \label{fig7}
\end{center}
\end{figure}

\subsection{N3 and NN of opposite signs}

If the sign of $t''$ is opposite to the sign of $t$, the behavior
of the Dirac points is very different. Thus for $t''+t<0$ the band
structure is not strongly modified (see Fig.~8). However at
$t''=-t$ a new quadratic point arises at the center of the BZ,
which splits into six classical Dirac points. With increasing
$t''$, these points evolve towards the same configuration as for large and positive values of $t''$.

\begin{figure}[htbp]
\begin{center}
\includegraphics[width=5in]{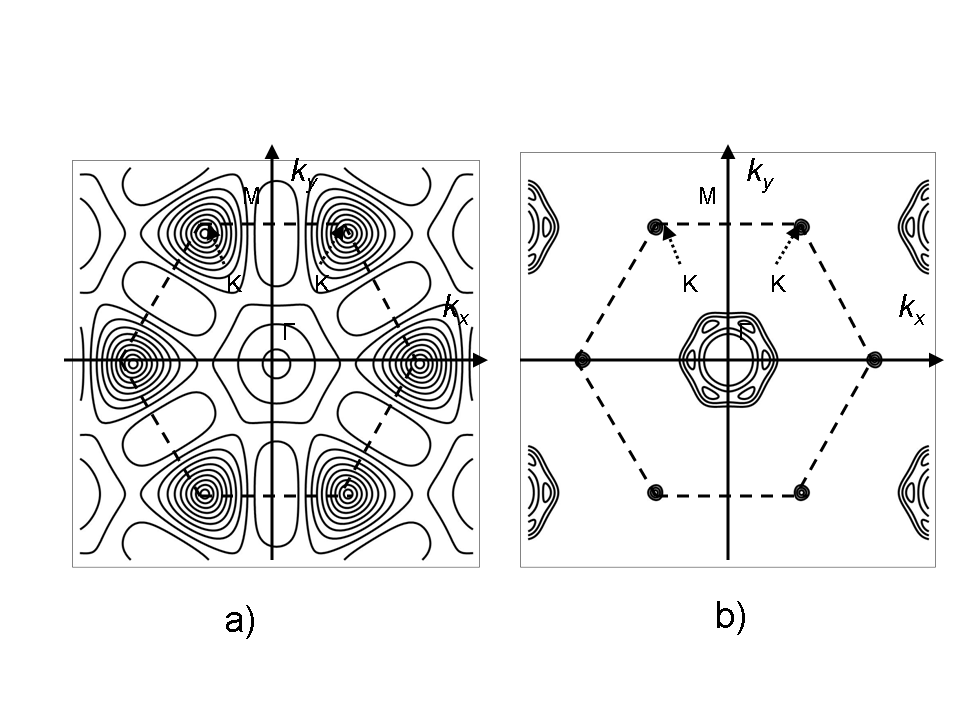}
\vspace{-0.3in} \caption{\small Graphene band structure for values
of $t''$ opposite in sign to $t$: a) Equal energy contours for
$t''=-0.5t$. The range of energies described here denotes energies
below the VH singularity. Note the band bending induced by the N3
coupling, which generates the extra contours around the M and
$\Gamma$ points. b)Low-energy contours for $t''=-1.4 t$. Note the
appearance of the extra six Dirac points close to $\Gamma$, the
center of the BZ.} \label{fig10}
\end{center}
\end{figure}

\section{Increasing the N2 coupling}

If we take into account $t'$, other effects also come into place, however these are not as dramatic as those of $t''$. The most important are an asymmetry between the valence and conduction bands, an
inclination of the Dirac cones, as well as a shift
of the energy of the Dirac points (effective doping). A few examples are given in Fig.~9.

\begin{figure}[htbp]
\begin{center}
\includegraphics[width=5in]{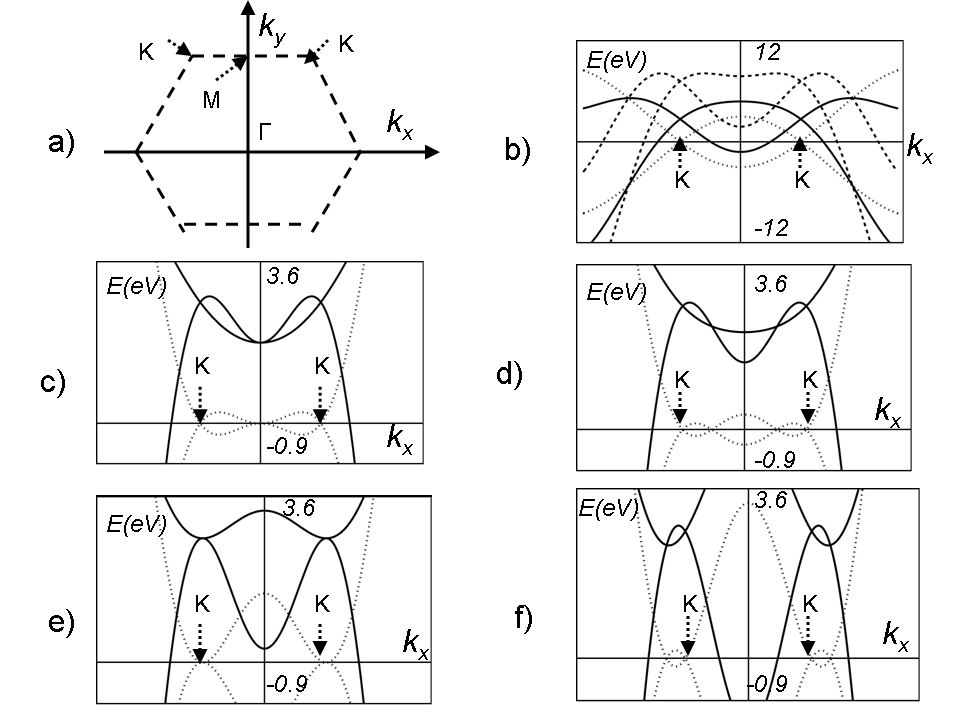}
\vspace{-0.01in} \caption{\small Graphene energy dispersion cuts
along the line KMK (as described in a) ) for b) $t''= 0$ and
$t'=0$ (dotted lines), $t''=0$ and $t'=0.3$ (dashed lines) and $t''=0$ and $t'=0.8 t$ (full lines).  In c)-f) we take c) $t''=t/3$ and $t'=0$ (dotted lines) and $t''=t/3$ and $t'=0.3t$ (full lines) d) $t''=0.4 t$ and $t'=0$ (dotted lines) and $t''=0.4t$ and $t'=0.3t$ (full lines) e)$t''=0.5 t$ and $t'=0$ (dotted lines) and $t''=0.5t$ and $t'=0.3t$ (full lines) f)$t''=0.7t$ and $t'=0$ (dotted lines) and $t''=0.7t$ and $t'=0.3t$ (full lines).} \label{fig8}
\end{center}
\end{figure}

\section{Density of states}
In the previous chapters we have focused on the energy dispersion. This chapter is devoted to the analysis of the density of states (DOS) of graphene, and of the evolution of the VH singularities in the presence of an N3 coupling. We calculate the
DOS by using the Hamiltonian in Eq.~(\ref{ht}), which we invert to
obtain the Green's function \be G(\omega,\k)=[\omega + i\delta
+H(\k)]^{-1} \ee The quantity $\delta$ is a phenomenological
inverse quasiparticle lifetime. The DOS is obtained from the
Green's function above by integrating it over the BZ. In Fig.~10
we plot the DOS for various ranges of parameters, and we can see
that the dynamics of the Dirac points is associated with changes
in the DOS.

\begin{figure}[htbp]
\begin{center}
\includegraphics[width=5in]{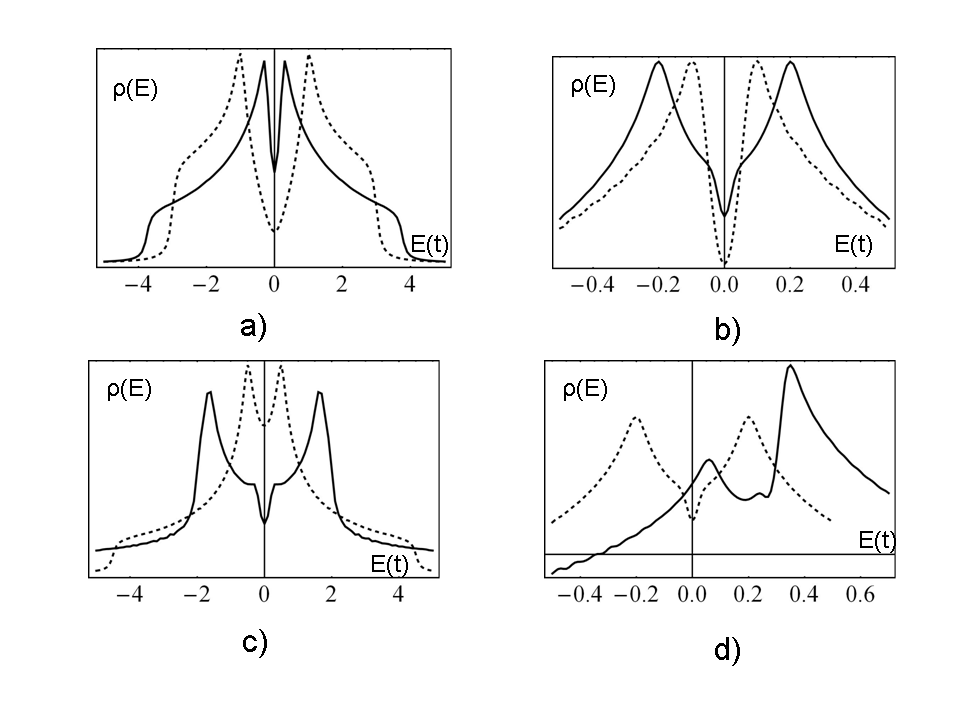}
\vspace{-0.01in} \caption{\small Graphene density of states for a)
$t''= 0$ (dashed) and $t''=t/4$ (full line), b)$t''=t/3$ (dashed
line) and $t''=0.4 t$ (full line) c) $t''=t/2$ (dashed line) and
$t''=0.7 t$ (full line). For a),b), and c) we set $t'=0$. In d) we
show the DOS for $t''=0.4 t$, and $t'=0$ (dotted line) and $t''=0.4 t$ and $t'=0.5t$ (full line).} \label{fig8}
\end{center}
\end{figure}

As expected, the low-energy DOS is indeed linear in the presence of ``typical'' linear
Dirac points in the spectrum, such as it is the case when $t''=0$, $1/3t < t''<
1/2 t$ and $t''>1/2 t$. The slight rounding in the figures is due
to the phenomenological inverse quasiparticle lifetime $\delta$
introduced for the convergence of the numerical calculation. In
Fig.~10 a) and d) this is of $0.1 t$, while in b) and c) it is of
$0.03t$. These values are larger than the physical estimates, but we chose them as such since smaller values of $\delta$ make the integrals more difficult to evaluate numerically. Moreover, the main effect of decreasing $\delta$ consists in a sharpening of the VH features. For example we have checked that the small bumps at
$t''=0.4t$ in Fig. 10 b) become indeed sharp divergences for values of $\delta$ smaller than $0.01 t$.

The contribution of a hybrid Dirac point to the low energy density of states should go as $\sqrt{|E|}$ \cite{jap}. Thus at  $t''=1/3t$ the density of states at low energy is expected to be a sum of a linear term (the contribution of the regular Dirac points) and of a square root term (the contribution of a hybrid point). In the presence of a quadratic
point at $t''=1/2t$ the density of states is expected to be
quadratic. This is consistent with the observations in Fig.~10.

We also note that for $1/3t < t'' <1/2 t$, as well as for $t''>1/2
t$ the VH peak splits into two, and in some range of parameters
(for example for $1/3t < t'' <1/2 t$), the energy of one of the two resulting peaks is greatly reduced (order of $0.1eV$). Thus it is conceivable that in this range of
parameters, the energies necessary to access the VH physics can
become accessible experimentally. A proposal to decrease the
energy of the VH singularities by rotating the successive graphene
planes by small angles was proposed in \cite{andrei}. Here we show
that this could be done as well by modifying the N3 coupling in
graphene or graphene-like materials. Thus we expect that in these materials more exotic physics such as charge density waves and superconductivity may also arise at smaller and more accessible energy scales.

\section{Conclusions and perspectives}
We have found a manner of modifying the number of
Dirac points and their positions in graphene-like materials,  by adjusting the second-nearest-neighbor and especially the third-nearest-neighbor hopping between atoms disposed in a honeycomb lattice. This
technique has advantages and disadvantages with respect to
previously proposed methods. The required N3 couplings are not too
large, of the order of $1/3$ NN (they are presently evaluated at up to $~0.15 $ NN \cite{rotenberg,tb1}). In this regime, the position of the Dirac points changes significantly even for small modifications of N3. Moreover, larger values of N3 yield a wide range of Dirac-point topologies and configurations: multiple linear points, hybrid points, quadratic points, etc. On the other hand, the displacement of the Dirac points presented here does not give rise to a gap opening.

For realizing experimentally such a system, the atomic structure should preserve the crystal symmetry (no preferred axis). Thus, some modifications of graphene via `internal' (lattice and chemical) factors such as modifying the doping, introducing various type of ad-atoms \cite{rotenberg}, as well as changing the nature of the substrate, may be more adequate than external factors such as applying pressure. For example, in recent
works it has been shown that a network of gold atoms can be
intercalated between the substrate and the first graphene layer on epitaxial graphene \cite{simon1}. In such a system spectroscopic features consistent with the existence of quasiparticles pockets at the M points have been observed \cite{simon2}, but no conclusive statement about the existence of new Dirac points at M could be drawn.

Other possibilities include focusing on multi-atom compounds exhibiting a triangular or hexagonal lattice, for example $Nb Se_2$ \cite{nbse2}. Furthermore, a hybrid bilayer system for which the atoms in the second layer are located above the centers of the hexagons in the first layer could have stronger N3 couplings, as these atoms may serve as `bridges' between opposing N3 neighbors in the first layer, while not affecting directly the NN and N2 couplings. One can also try creating a system with large N3 using cold atoms.

We have also shown that modifying the configuration of Dirac points corresponds to an adjustment of the VH singularities. In particular the N3 coupling can significantly shift the energy of the VH singularities towards the experimentally-accessible range. As also noted in \cite{andrei}, this is is an important step
towards exploring the exotic physics oftentimes associated with the VH singularities such as charge density waves, and superconducting instabilities. From a more theoretical perspective it would be interesting to see if our analysis can be extended to cuprates, which are also known to exhibit linearly dispersing quasiparticles, as well as VH singularities.

\section{Acknowledgements}
We would like to thank Gilles Montambaux, Jean-No\"{e}l Fuchs, Frederic Piechon and Mark Goerbig for interesting discussions. CB would like to acknowledge funding from the French ANR (Agence Nationale de Recherche), under the P'NANO program, reference NANOSIM-GRAPHENE.

\end{document}